\documentclass[a4paper]{jpconf}
\usepackage{graphicx}
\usepackage{amsmath,amsfonts,amssymb}

\def\ov{\overline}

\def\mbb{\mathbb}

\begin{document}
\title{Searching for axions and ALPs from string theory}

\author{Andreas Ringwald}

\address{Deutsches Elektronen-Synchrotron DESY, Notkestr. 85, D-22607 Hamburg, Germany}

\ead{andreas.ringwald@desy.de}

\begin{abstract}
We review searches for closed string axions and axion-like particles (ALPs) in IIB string
flux compactifications.  
For natural values of the background fluxes and TeV scale gravitino mass, the moduli stabilisation mechanism of the 
LARGE Volume Scenario predicts the existence of a QCD axion candidate with intermediate 
scale decay constant, $f_a\sim 10^{9\div 12}$\,GeV, associated with 
the small cycles wrapped by the branes hosting the visible sector, plus a nearly massless and nearly decoupled
ALP associated with the LARGE cycle. 
In setups where the visible sector branes are wrapping more than the minimum number of two intersecting cycles, 
there are more ALPs which have approximately the same decay constant and coupling to the photon as the 
QCD axion candidate, but which are exponentially lighter. There are exciting phenomenological opportunities 
to search for these axions and ALPs in the near future. For $f_a\sim 10^{11\div 12}$~GeV, 
the QCD axion can be the dominant part of dark matter and be detected in haloscopes exploiting microwave cavities. 
For $f_a\sim 10^{9\div 10}$~GeV, the additional ALPs could explain astrophysical anomalies and
be searched for in the upcoming generation of helioscopes and light-shining-through-a-wall
experiments.
\end{abstract}

\section{Introduction}

The QCD axion arises~\cite{Weinberg:1977ma,Wilczek:1977pj} in the course of the arguably most plausible solution of the strong CP puzzle~\cite{Peccei:1977hh}, that is the non-observation of a $\theta$-angle term in QCD. In this context, 
the axion field $a$ is introduced as a dynamical $\theta$-angle term, enjoying a shift symmetry,
$a\to a + {\rm constant}$, 
broken only by anomalous CP-violating couplings to gauge fields. Correspondingly, its most general 
low-energy effective Lagrangian below the weak scale has the form~\cite{Georgi:1986df},
\begin{eqnarray}
\mathcal{L} &= \frac{1}{2}\, \partial_\mu a\, \partial^\mu a
- \frac{g_{3}^2}{32\pi^2} \left(\bar{\theta}  +  \frac{a}{f_{a}} 
\right) F_{3,\mu\nu}^b \tilde{F}_3^{b,\mu\nu}  - \frac{e^2}{32\pi^2} C_{a\gamma} \frac{a}{f_{a}} \, F^{\rm em}_{\mu\nu} \tilde{F}^{\mu\nu}_{\rm em} 
\nonumber\\
&  + \sum_\Psi \bigg[\frac{1}{2} (\tilde{X}_{\psi_R} + \tilde{X}_{\psi_L}) \ov{\Psi} \gamma^\mu\gamma_5   \Psi  + \frac{1}{2} (\tilde{X}_{\psi_R} - \tilde{X}_{\psi_L}) \ov{\Psi} \gamma^\mu \Psi \bigg] \frac{\partial_\mu a}{f_{a}}
\, ,
\label{axion_leff}
\end{eqnarray}
where $F_{3,\mu\nu}^b$ is the $b$th component of the gluon field strength tensor, $\tilde{F}_3^{b,\mu\nu} = \frac{1}{2} \epsilon^{\mu\nu\rho \sigma}F_{3,\rho\sigma}^b $ its dual, $g_3$ is the strong coupling, 
$F^{\rm em}_{\mu\nu}$ is the electromagnetic field strength, 
$\Psi$ denotes standard model matter fields, and $f_a$ is the axion decay constant.
The $\theta$-term in the QCD Lagrangian can then be
eliminated by absorbing it into the axion field, $a=\bar{a} - \bar{\theta}
f_a$. Moreover, the topological charge density $\propto \langle
F^b_{3,\mu\nu} {\tilde F}_3^{b,\mu\nu} \rangle \neq 0$, induced
by topological fluctuations of the gluon fields such as QCD
instantons, provides a nontrivial potential for the axion field $\bar{a}$
which is minimized at zero expectation value, $\langle \bar{a}\rangle =0$, 
thus wiping out strong CP violation, and giving the
fluctuating field around this minimum, the QCD axion, a mass 
\begin{eqnarray}
m_a =
         \frac{m_\pi f_\pi}{f_a}\frac{\sqrt{m_u m_d}}{m_u+m_d}\simeq { 0.6\,  {\rm meV}}
         \times
         \left(
         \frac{10^{10}\, {\rm GeV}}{f_a}\right), 
         \label{axionmass}
\end{eqnarray}
in terms of the light ($u,d$) quark masses,
the pion mass $m_\pi$ and the pion decay constant $f_\pi$~\cite{Weinberg:1977ma},
For large axion decay constant $f_a$, we see that the axion is a very weakly interacting (cf. Eq.~(\ref{axion_leff})) 
slim particle~\cite{Kim:1979if,Dine:1981rt,Shifman:1979if,Zhitnitsky:1980tq}. In particular, its coupling to 
photons~\cite{Bardeen:1977bd,Kaplan:1985dv,Srednicki:1985xd}
and electrons, 
\begin{eqnarray}
\mathcal{L} \supset -  \frac{\alpha}{2\pi f_a} \left( C_{a\gamma} - {\frac{2}{3}\,\frac{m_u+4 m_d}{m_u+m_d}     }\right) \,\frac{a}{4} \,F^{\rm em}_{\mu\nu} \tilde{F}^{\mu\nu}_{\rm em} + \frac{C_{ae}}{2 f_{a}} \,\bar{e} \gamma^\mu\gamma_5 e \partial_\mu a\,,
\label{EQ:DEFCAGG}
\end{eqnarray}
are very much suppressed, e.g. 
\begin{eqnarray}
        { g_{a\gamma}} \equiv \frac{\alpha}{2\pi f_a}
\underbrace{\left( C_{a\gamma} - {\frac{2}{3}\,\frac{m_u+4 m_d}{m_u+m_d}     }\right)}_{\mathcal{C}_{a\gamma}}
\sim 10^{-13}\ {\rm GeV}^{-1}          \left(
         \frac{10^{10}\, {\rm GeV}}{f_a}\right) \mathcal{C}_{a\gamma}.
         \label{axionphotoncoupling}
\end{eqnarray}

\begin{center}
\begin{figure}[h]
\begin{center}
\includegraphics[width=0.7\textwidth]{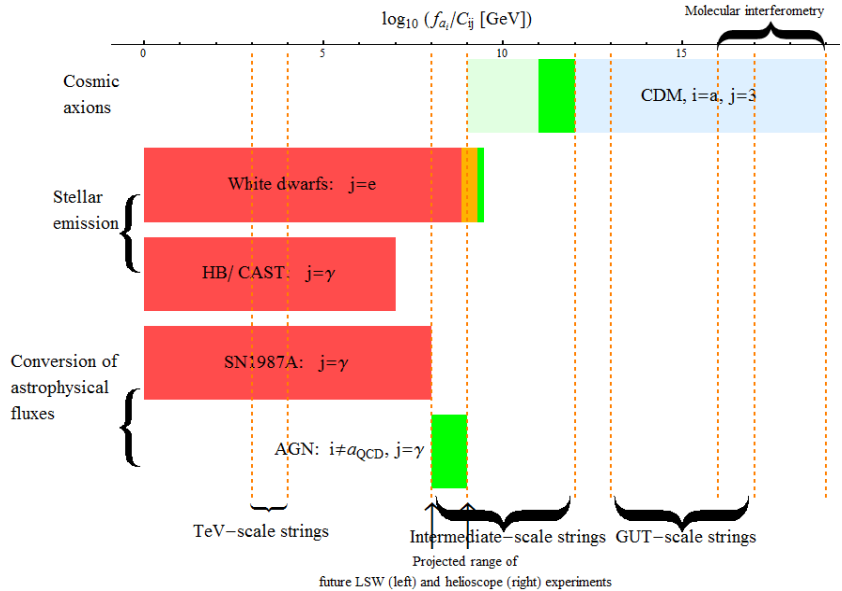}
\end{center}
\caption{A summary of constraints on and hints for the couplings $f_{a_i}/C_{ij}$ of axion-like particles $a_i$ to 
standard model particles $j$~\cite{Cicoli:2012sz}.
Red regions are excluded,
and the orange region would be excluded by red giants but is compatible with the hints from white dwarfs. 
The green regions from top to bottom correspond respectively to the classic `axion dark matter window', hints of an axion from
white dwarf cooling and transparency of the Universe to very high energy gamma rays. The blue region would be excluded by dark matter overproduction in the absence of a dilution mechanism or tuning of the initial misalignment angle.
}
\label{FIG:Constraints}
\end{figure}
\end{center}

Laboratory experiments as well as astrophysics, in particular the non-observation of solar axions by 
the helioscope CAST and the non-observation of drastic
energy losses in horizontal branch stars or white dwarfs, constrain the axion decay constant, 
divided by the appropriate dimensionless coupling constants, to a scale much above the weak scale
(see also Fig.~\ref{FIG:Constraints}), 
\begin{eqnarray}
\frac{f_{a}}{\mathcal{C}_{a\gamma}} > 10^7\ {\rm GeV} \Leftrightarrow 
g_{a\gamma}< 10^{-10}\ {\rm GeV}^{-1}\,,
\quad \frac{f_{a}}{C_{ae}} > 10^9\ {\rm GeV}.
\end{eqnarray}

Intriguingly, for an even higher decay constant, $f_a\gtrsim 10^{11}$~GeV, the QCD axion can contribute 
significantly to cold dark matter (CDM), being non-thermally produced in the early universe via initial 
misalignment of the axion field, resulting in coherent field oscillations 
corresponding to a condensate of non-relativistic axions~\cite{Preskill:1982cy,Abbott:1982af,Dine:1982ah}. 
In fact, assuming that the reheating temperature after inflation
is below $f_a$ and that there is no dilution by, e.g., late decays of particles beyond the standard model,   
the expected cosmic mass fraction in QCD axion CDM is
\begin{equation}
\Omega_a h^2\approx  0.71 \times \left( \frac{f_a}{10^{12}\ \rm GeV} \right)^{7/6} \left( \frac{\Theta_a}{\pi} \right)^2,
\label{eq:omegaqcdaxion}
\end{equation}
where $\Theta_a$ is the initial misalignment angle.  

Therefore, the QCD axion is necessarily associated with a very high energy scale 
and so it is natural to search for it in ultra-violet completions of the standard model 
such as string theory.
Indeed, it has long been known that the low-energy effective field theory of string compactifications predicts
natural candidates for the QCD axion~\cite{Witten:1984dg,Conlon:2006tq,Svrcek:2006yi,Conlon:2006ur,Choi:2006qj},
often even an `axiverse'~\cite{Arvanitaki:2009fg}, containing
many additional light axion-like particles (ALPs) whose masses are logarithmically hierarchical.   
But only very recently explicit moduli-stabilised string theoretic examples 
with a viable QCD axion candidate and possibly additional light ALPs have been 
constructed (cf. Sec.~\ref{sec:iibstringaxiverse}), with decay constants ranging from the GUT scale, $f_a\sim 10^{16}$~GeV~\cite{Acharya:2010zx} 
down to the intermediate scale, $f_a\sim 10^{9\div 12}$~GeV~\cite{Cicoli:2012sz} -- the latter 
offering exciting opportunities to detect effects of axions and ALPs in astrophysics and in the upcoming generation of 
axion experiments (cf. Sec.~\ref{sec:opportunities}). 
  
\section{\label{sec:iibstringaxiverse}Axions and ALPs in IIB string flux compactifications}

String theory requires the existence of six extra space dimensions, which appear to be
unobservable because they are supposed to be compact and of very small size. 
String phenomenology is then the attempt to make contact between the perturbative ten dimensional (10D) 
effective field theories (EFTs) describing the massless degrees of freedom of string theory at very high energies 
-- say, the heterotic or type II (with $D$-branes) EFT -- and the low energy physics in our 4D real world. 
In fact, different 4D low energy EFTs emerge depending on 
the EFT to start with in 10D. One of the fundamental tasks of string phenomenology is 
to find a compactification whose low energy EFT reproduces 
(a suitable extension of) the standard model. 
Along this way it is also mandatory to understand moduli stabilisation: expectation values 
of the moduli fields, which parameterise the shape and size of the extra dimensions,  
determine many parameters of the low energy EFT, such as gauge and  
Yukawa couplings. Often, one needs to incorporate quantum corrections in order to fix 
many of these expectation values and to give their associated particle
excitations a non-zero mass.  

Moduli stabilisation is best understood in type IIB string flux compactifications, on which we 
concentrate in the following. They provide explicit and well motivated realisations of brane world scenarios:  
the standard model is supposed to live on a stack of space-time filling branes
wrapping cycles in the compact dimensions, while gravity propagates in the bulk, leading 
to a string scale $M_s\sim M_P/\sqrt{\mathcal{V}}$ possibly much smaller than the Planck scale $M_P$, 
at the expense of a large compactification volume $\mathcal{V}\gg 1$ (in units of the string length).  
Importantly, axion-like fields emerge in those compactifications inevitably 
as Kaluza-Klein zero modes of ten dimensional form fields, as we will review in the following 
subsection.

\subsection{\label{sec:treelevel} Tree-level candidates for axions and ALPs}

We will consider here type IIB flux compactifications on Calabi-Yau orientifolds $X$ in the presence 
of space-time filling $D7$ branes and $O7$ planes. Below the Kaluza-Klein scale $M_{\rm KK}$,  
they lead to a low-energy $\mathcal{N} = 1$ (supergravity) EFT in 4D.
Their 4D closed string moduli, comprised by the axio-dilaton $S=e^{-\phi }+i C_0$,
the complex structure moduli $U_{\alpha}$, $\alpha=1,...,h_-^{2,1}(X)$, and the K\"ahler moduli,
\begin{eqnarray}
T_i =\tau_i + i\, c_i,\quad\tau_i 
=\textrm{Vol}(D_i),\quad c_i =\int_{D_i }C_4,\quad i =1,...,h^{1,1}\,, 
\end{eqnarray}
are obtained via the Kaluza-Klein reduction of the massless bosonic fields of the original 10D theory -- the latter including, in the Ramond-Ramond sector, the forms $C_{0}$ and $C_4$, and, in 
the Neveu Schwarz-Neveu Schwarz sector, the dilaton $\phi$.  
Importantly, the number of K\"ahler moduli $T_i$ is determined by the topology of $X$, namely the number of 
inequivalent four-cycles $D_i$ of $X$ (we have specialised for convenience to 
orientifold projections such that $h^{1,1}_- = 0 \Rightarrow h^{1,1}_+ = h^{1,1}$).  
Their imaginary parts $c_i$ have all the properties of axion-like fields, as we will see next. 

As already mentioned, (a suitable extension of) the standard model is realised in such 
compactifications through (stacks of) space-time filling $D7$-branes wrapping some of the 
four-cycles $D_i$. At low energies, the dynamics of a $D7$-brane reduces to a $U(1)$ gauge theory that lives on its 
eight-dimensional world-volume. Moreover, $D7$-branes can be magnetised by turning 
on internal magnetic fluxes $\mathcal F$. Taking all these effects into account,  
the 4D low energy effective Lagrangian of the axion-like fields $c_i$, including the brane-localised 
$U(1)$ gauge bosons $A_i$, is obtained from the Kaluza-Klein reduction of the $D7$-brane action   
as follows~\cite{Grimm:2004uq,Jockers:2004yj}:
\begin{align}
\mathcal{L} &\supset  
-  \left( d c_\alpha + \frac{M_P}{\pi} A_i q_{i\alpha}\right) \frac{\mathcal{K}_{\alpha\beta}}{8}
\wedge \star \left( d c_\beta + \frac{M_P}{\pi} A_j q_{j\beta}\right) 
+\frac{1}{4\pi M_P}  r^{i\alpha} c_\alpha \tr(F\wedge F)\nonumber\\
&
+ \frac{M_P^2}{2(2\pi)^2}  A_i A_j q_{i\alpha}  \mathcal{K}_{\alpha\beta} q_{j\beta}
 - \frac{r^{i\alpha}\tau_\alpha}{4 \pi M_P} \tr(F_i \wedge \star F_i)  ,
\label{EQ:BIGAXIONCOUPLINGS}
\end{align}
where $M_P=(8\pi G_N)^{-1/2} \simeq 2.4 \cdot 10^{18}$ GeV is the reduced Planck mass,
while the various other quantities in Eq.~\eqref{EQ:BIGAXIONCOUPLINGS} are defined as follows: 

\begin{itemize}

\item 
The K\"ahler metric $K_{\alpha\beta}$, describing the kinetic mixing of the axion-like fields,
is obtained via $\mathcal{K}_{\alpha\beta} \equiv \frac{\partial^2 K}{\partial\tau_\alpha\partial\tau_\beta}$
from the K\"ahler potential, the latter taking, at tree-level, the following form:
\begin{equation}
  K_{\rm tree}=-2\ln \mathcal{V} -\ln
  \left(S+\bar{S}\right) -\ln \left( -i\int\limits_X \Omega \wedge
  \bar{\Omega}\right)\,. 
  \label{eqtree}
\end{equation}
It depends implicitly on the complex structure moduli via the holomorphic (3,0)-form $\Omega$ and on the K\"ahler moduli via the Calabi-Yau volume $\mathcal{V}$, measured by an Einstein frame
metric $g^{\scriptscriptstyle E}_{\mu \nu} = e^{-\phi/2} \,
g^s_{\mu \nu}$,  and expressed in units of the string length $\ell_s = 2\pi \sqrt{\alpha'}$, in terms of its tension $\alpha'$,
\begin{equation}
  \mathcal{V} = \frac 16 \int_X J\wedge J\wedge
  J = \frac 16 \, k_{\alpha\beta\gamma} t^\alpha t^\beta t^\gamma\,. \label{Vol}
\end{equation}
Here, the K\"ahler form $J$ has been expanded, 
in a basis $\{ \hat{D}_\alpha \}_{\alpha=1}^{h^{1,1}}$ of $H^{1,1}(X,\mbb{Z})$ of two-forms which are Poincar\'e
dual to $D_\alpha$,  
as $J = t^\alpha \hat{D}_\alpha$, and we denoted the triple intersection numbers of $X$ by $k_{\alpha\beta\gamma}$.
The volume, the K\"ahler form and ultimately the K\"ahler metric can then be obtained as a function of the $\tau_\alpha$ by inverting the following relations:
\begin{equation}
 \tau_\alpha  = \frac 12\int_X \hat{D}_\alpha\wedge J\wedge J
  =\frac{\partial \mathcal{V}}{\partial t^\alpha}
  =\frac 12\, k_{\alpha\beta\gamma}\, t^\beta\, t^\gamma\,. \label{TauDef}
\end{equation}
The string scale $M_s = 1/\ell_s$ is obtained from the dimensional reduction of the
IIB supergravity action,
\begin{equation}
 M_s = M_P / \sqrt{4\pi\mathcal{V}}.
\label{eq:stringscale}
\end{equation} 

\item 
The couplings $r^{i\alpha}$ 
appearing in the gauge kinetic term and in the axionic couplings to gauge bosons, 
the respective last terms in both lines of Eq.~\eqref{EQ:BIGAXIONCOUPLINGS}, 
are the expansion coefficients of the two form $\hat{D}_i = r^{i\alpha} \omega_\alpha$,  
\begin{equation}
r^{i\alpha} \equiv \,\ell_s^{-4} \int_{D_i} \tilde{\omega}^\alpha =  \ell_s^{-4} 
\int \hat{D}_i \wedge \tilde{\omega}^\alpha,\quad \alpha=1,...,h^{1,1},
\end{equation}
where the basic forms satisfy $\ell_s^{-4} \int \omega_\beta \wedge \tilde{\omega}^\alpha = \delta_\eta^\alpha$.
The gauge coupling can be inferred from the gauge kinetic term as
\begin{equation}
\frac{1}{g_i^2} = \frac{r^{i\alpha} \tau_\alpha}{2\pi M_P}\times
\left\{ \begin{array}{ll} 1 & U(1) \\ 1/2 & SU(N)\end{array} \right..
\end{equation}

\item 
The couplings $q_{i\alpha}$ appearing in the (St\"uckelberg) mass terms for the $U(1)$ gauge fields $A_i$ 
in Eq.~\eqref{EQ:BIGAXIONCOUPLINGS} are given by
\begin{equation}
q_{i\alpha} \equiv \,\ell_s^{-2} \int_{D_i} \omega_\alpha \wedge \frac{\mathcal{F}}{2\pi}  = 
\ell_s^{-4} \int_{D_i} \omega_\alpha \wedge \ell_s^2\mathcal{F}, 
\end{equation}
where $\mathcal{F}$ is the gauge flux on $D_i$. Axions $c_\alpha$ experiencing such a coupling     
disappear from the low energy EFT because they are eaten by the corresponding $U(1)$ gauge 
boson~\cite{Goodsell:2009xc,Cicoli:2011yh}.

\end{itemize}

Thus, it appears from Eq.~\eqref{EQ:BIGAXIONCOUPLINGS}, that IIB string flux compactifications 
have potentially many, $h^{1,1}-d$, axion candidates $c_i$, where $d$ is the number of $U(1)$ bosons
$A_i$ which get a St\"uckelberg mass by eating the associated axions. However, before reaching this
conclusion one has
to consider perturbative and non-perturbative corrections to this tree-level result. In fact, such 
corrections are necessarily to be taken into account in the course of the stabilisation 
of the associated K\"ahler moduli $\tau_i$.  
Importantly, the mechanisms to fix the $\tau_i$ may also generate large masses 
for the corresponding axions $c_i$~\cite{Banks:2003es,Donoghue:2003vs}, as we will see next.    

\subsection{Axions and ALPs in moduli stabilised IIB string flux compactifications}

In IIB string flux compactifications, the tree-level superpotential~\cite{GVW}
\begin{equation}
  W_{\rm tree}=\int\limits_X G_3 \wedge \Omega \,.
  \label{Wtree}
\end{equation}
which is generated by turning on background fluxes of the form $G_3 = F_3 +iS
H_3$, where $F_3= dC_2$ and $H_3=dB_2$, does not depend on the K\"ahler moduli,
but on the dilaton $C$ and the complex structure moduli $U_\alpha$.  
This implies that the dilaton and the complex structure moduli can be fixed at
tree-level by imposing vanishing F-term conditions~\cite{gkp}. By appropriate 
tuning of the internal fluxes, one can always fix the dilaton such that the 
string coupling, 
$g_s=1/{\rm Re}(S)$,  
is in the perturbative regime. Further 
effects from fixing $S$ and $U$ are then parametrised by the flux-dependent constant 
$W_0=\langle W_{\rm tree}\rangle$ 
and by an overall factor in the F-term scalar potential 
arising from the $S$ and $U$ dependent part in the tree-level K\"ahler potential 
\eqref{eqtree}.    

The K\"ahler moduli $\tau_i$, however, remain precisely massless at leading semiclassical order,
because of the no-scale structure of $K_{\rm tree}$. 
Their stabilisation requires taking into acount perturbative (p) and nonperturbative (np)  
$\alpha'$ and $g_s$ corrections to the tree-level result, 
\begin{equation}
W=W_{\rm tree} + \delta W_{\rm np}, \hspace{6ex} K = K_{\rm tree} + \delta K_{\rm p} + \delta K_{\rm np},
\end{equation}
eventually leading also to non-trivial potentials, i.e. masses, for the axions 
$c_i$ associated with the scalars $\tau_i$. These masses can arise, however, only via the 
non-perturbative corrections to the superpotential, $\delta W_{\rm np}$, and to the 
K\"ahler potential, $\delta K_{\rm np}$: 
only those   
can break the shift symmetry of the axions $c_i$. 

The respective order of magnitude of the perturbative versus the nonperturbative
corrections to the scalar potential is set by $W_0$. 
The main mechanisms proposed for $\tau$ moduli stabilisation and their consequences 
for the physics of their associated  $c$ axions can be characterised as follows:

\begin{itemize}

\item 
The first scenario of K\"ahler moduli stabilisation in IIB string compactifications neglected the corrections to the
K\"ahler potential and 
considered only the non-perturbative corrections to the superpotential~\cite{kklt},  
\begin{equation}
  \delta W_{\rm np} = \sum\limits_{i=1}^{h^{1,1}} A_i(S,U)\, e^{- a_i T_i}\,,
\label{eq:Wnp}
\end{equation}
which arises by $ED3$ instantons (in which case $a_i=2\pi $) 
or by stacks of $D7$ branes supporting a condensing gauge theory
(for which $a_i=6\pi /b_0$ with $b_0$ being the coefficient of the one-loop beta function),
wrapping the four-cycles $D_i$. 
The threshold effects $A_i$ can be considered as $\mathcal{O}(1)$ constants since
they depend on the complex structure moduli which are flux-stabilised at tree-level. 
The four-cycle volumina are then fixed at 
\begin{equation}
\tau_i \sim \frac{1}{a_i}\ln\left( \frac{W_0}{A_i}\right).
\end{equation}
Thus, $W_0$ has to be fine-tuned to extremely small values, $W_0\ll 1$, in order 
that the volume is large, $\mathcal{V}\sim \tau_i^{3/2}\gg 1$, the latter being a prerequisite of 
the underlying supergravity approximation. Therefore, in these scenarios it is extremely hard
to lower the string scale \eqref{eq:stringscale} much below the Planck scale. Moreover, since the axionic shift
symmetry $c_i\to c_i+{\rm constant}$ of all axions is broken non-perturbatively by Eq.~\eqref{eq:Wnp}, all the 
axion candidates get a
large mass of the order of the mass of the particle excitations of the associated K\"ahler moduli 
(the so-called ``saxions"), 
$$m_{c_i}\sim m_{\tau_i}\sim a_i W_0 M_P/\mathcal{V},$$ 
leaving no candidate for a QCD axion~\cite{Conlon:2006tq}, let alone a light ALP.

\item This can be avoided if the non-pertubative effects arise from 
wrapping an {\it ample} four-cycle,  
\begin{equation}
D_{\rm am}=\sum_{i=1}^{h^{1,1}} \lambda_i D_i,\ {\rm with\ }\ \lambda_i >0\  \forall\,i=1,...,h^{1,1},
\end{equation}
in terms of a basis $\{D_i\}$ of $H_4(X,\mathbb{Z})$,  
resulting in a contribution to the superpotential of the form
\begin{equation}
\delta W_{\rm np} = A\,e^{-\,a\, T_{\rm am}}= A\,e^{-\,a \sum_{i=1}^{h^{1,1}} \lambda_i T_i}.
\end{equation}
In this case, by fine-tuning $W_0 \sim A\,e^{-\,a\, T_{\rm am}} \ll 1$,
this single non-perturbative effect can generate a minimum for all the K\"ahler moduli $\tau_i$ within the regime of validity
of the EFT~\cite{Bobkov:2010rf}. 
However it can lift only one axion corresponding to the imaginary part of the ample divisor modulus: $c_{\rm am}={\rm Im}(T_{\rm am})$.
All the remaining $h^{1,1}-1$ axions are massless at leading order
and develop a potential only via tiny higher order instanton effects of the form~\cite{Acharya:2010zx}:
\begin{equation}
\delta W_{\rm np}=A\,e^{-\,a \,T_{\rm am}} + \sum_{i=1}^{h^{1,1}-1} A_i\,e^{-\, n_i a_i T_i},
\label{higherNP}
\end{equation}
where $T_i$ is a combination of moduli orthogonal to $T_{\rm am}$ $\forall\,i=1,...,h^{1,1}-1$.

Thus, this moduli stabilisation scenario gives rise to an axiverse~\cite{Arvanitaki:2009fg}: 
possibly many, $h^{1,1}-1$, axions which acquire a mass spectrum which is logarithmically hierarchical.
This number may still be diminished in case that some of the axions are eaten by the St\"uckelberg mechanism
to provide masses to brane-localised $U(1)$ gauge bosons.   

Two main concerns regarding the microscopic realisation of this scenario
are related to the difficulty to find an ample divisor which is rigid (and so definitely receives non-perturbative effects),
and the possibility to choose gauge fluxes that avoid chiral intersections between the instanton and the visible sector.
Moreover, it should be noted that in this case again very large volumina are not possible: the string 
scale \eqref{eq:stringscale} is expected not much below the Planck scale, of order the GUT scale. 

\item The latter difficulties are avoided in the LARGE volume scenario (LVS), which realizes the 
possibility of exponentially large volumina for generic values of $W_0 \sim\mathcal{O}(1)$~\cite{Balasubramanian:2005zx}  
and allows for the construction of explicit globally and locally 
consistent Calabi-Yau examples with magnetised D7-branes, realising MSSM or GUT like chiral extensions
of the standard model~\cite{Cicoli:2011qg,Cicoli:pascos}. 

The LVS requires the existence of a single {\it del Pezzo} four-cycle $\tau_{\rm dP}$ which guarantees 
the emergence of a non-perturbative contribution to the superpotential  
\begin{equation}
\delta W_{\rm np}= A\,e^{-\,a\,T_{\rm dP}}
\end{equation} 
via an $ED3$ instanton or gaugino condensation. 
This effect fixes $\tau_{\rm dP}$ at a small size and gives the corresponding saxion and 
axion a large mass~\cite{LVSspectrum}, 
\begin{equation}
m_{\tau_{dP}} \sim m_{c_{\rm dP}}  \sim 
\frac{W_0 \sqrt{\ln \mathcal{V}}}{\mathcal{V}}\,M_P\,.
\end{equation}
All the other $\tau$ moduli are stabilised perturbatively by $\alpha'$ or $g_s$ effects or by D-terms 
arising from magnetised branes. 
The exponential large volume emerges from an interplay between the non-perturbative contribution 
associated with del Pezzo cycle and the leading $\alpha'$ correction~\cite{alphaprime}:
\begin{equation}
\delta K_{\rm p} 
\simeq 
-\frac{\zeta}{g_s^{3/2}\mathcal{V}}\,,\ {\rm with}\ \zeta \propto (h^{1,2}-h^{1,1})\,,  
\end{equation}
which yields, for $h^{1,2}>h^{1,1}>1$ (i.e. negative Euler number), a supersymmetry-breaking 
anti de Sitter (AdS) minimum at exponentially large volume~\cite{Balasubramanian:2005zx}:
\begin{equation}
\mathcal{V} \sim W_0 \,e^{\,a\,\tau_{\rm dP}}\,.
\end{equation}

Hence only one axion, $c_{\rm dP}={\rm Im}(T_{\rm dP})$, becomes heavy whereas all the other ones
(except those eaten up by anomalous $U(1)$s, whose scalar partners are fixed by the above mentioned D-terms) 
remain light and develop a potential via subleading higher order instanton effects.

\begin{figure}[h]
\includegraphics[width=0.35\textwidth]{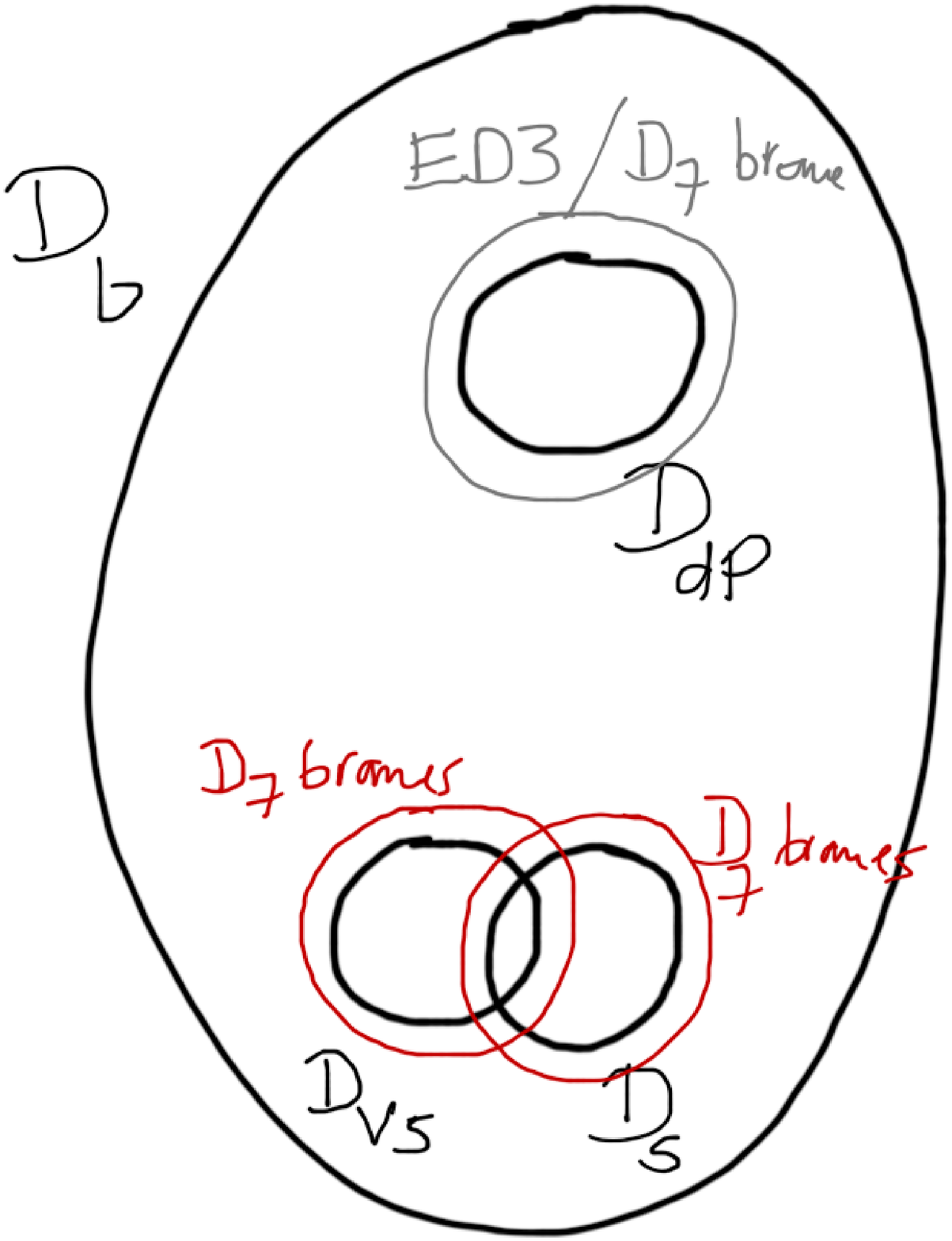}\hspace{2pc}%
\begin{minipage}[b]{0.5\textwidth}\caption{\label{fig:lvs}Schematic picture of the simplest
Swiss-cheese setup of a compactification in the LARGE volume scenario: a big four-cycle (=divisor) $D_b$ with exponentially large 
volume $\tau_b\sim \mathcal{V}^{2/3}$; a del Pezzo divisor $D_{\rm dP}$ supporting the leading non-perturbative
effect ($ED3$ instanton or gaugino condensation of a pure Yang-Mills theory described by a stack of $D7$ branes); 
a rigid divisor $D_{\rm vs}$ supporting the stack of $D7$ branes describing the visible sector; 
a rigid divisor $D_{\rm s}$ intersecting with $D_{\rm vs}$ and supporting a stack of magnetised 
$D7$ branes for D-term stabilisation.}
\end{minipage}
\end{figure}

The simplest version of a standard LVS is build upon a Swiss-cheese Calabi-Yau 
three-fold with volume given by \cite{ExplicitSC}:
\begin{equation}
\mathcal{V}=\alpha\left(\tau_b^{3/2}-\sum_{i=1}^{h^{1,1}-1} \gamma_i \tau_i^{3/2}\right).
\end{equation}
It is dominated by the exponentially large volume $\tau_b$ of one of the four-cycles, $D_b$,  
while all the other, $h^{1,1}-1$, four-cycles are small. 

Importantly, the realisation of an LVS requires $h^{1,1}\geq 4$ (cf. Fig.~\ref{fig:lvs}): a single del Pezzo divisor  
$D_{\rm dP}$ to support the non-perturbative effects, one big four-cycle $\tau_{\rm b}$ to parametrise the
large volume, $\mathcal{V}\sim \tau_{\rm b}^{3/2}$, one small rigid four-cycle $D_{\rm vs}$ to support the stack 
of space-time filling magnetised D7-branes corresponding to (a suitable extension of) the MSSM and 
intersecting with another small cycle $D_{\rm s}$ whose brane setup is there to provide 
D-terms which stabilise $\tau_{\rm vs}$ and thus the visible sector gauge coupling $g_{\rm vs}^{-2}\sim \tau_{\rm vs}$.
Two of the $h^{1,1}\geq 4$ axion-like fields disappear from the low energy spectrum, however: 
$c_{\rm dP}$, as explained above, and $c_{\rm s}$, which is eaten by the St\"uckelberg mechanism. 
Thus, realisations of the LVS axiverse involve at least two light axions: one 
QCD axion candidate plus one ALP~\cite{Cicoli:2012sz}. More generic models for $h^{1,1}$ very large will
include an arbitrarily large number of ALPs.

The main scales in the model are:
\begin{eqnarray}
M_s &= &\frac{M_P}{\sqrt{4\pi {\mathcal V}}} \sim 10^{10}\,{\rm GeV},  
m_{\tau_{\rm s}}\sim \frac{M_P}{\mathcal V^{1/2}} \sim 10^{10}\,{\rm GeV}, 
M_{\rm KK} \sim \frac{M_P}{{\mathcal V}^{2/3}} \sim 10^{9}\,{\rm GeV},\nonumber\\
m_{\tau_{\rm dP}}&\sim & \sqrt{g_s}\,W_0\frac{M_P}{\mathcal V}\ln{\mathcal V}\sim 30\,{\rm TeV}, 
m_{\rm soft}\sim m_{3/2}\sim  \sqrt{g_s}\,W_0 \frac{M_P}{\mathcal V}\sim 1\,{\rm TeV}, \\
m_{\tau_{\rm vs}}&\sim &\alpha_{\rm vs} m_{3/2}\sim 40\,{\rm GeV}, 
m_{\tau_b}\sim \frac{M_P}{\mathcal V^{3/2}} \sim 0.1\,{\rm MeV}.\nonumber
\end{eqnarray}
The numerical values have been given for generic values of the underlying parameters,  
$g_s\sim 0.1, W_0\sim 1$, and for a volume ${\mathcal V} \sim 10^{14}$, demonstrating that, 
for an intermediate string scale, the LVS naturally realises TeV-scale SUSY. 

\item Finally, the contributions from both $\delta K_{p}$ and non-perturbative terms in the superpotential $\delta W_{np}$ from gaugino condensation on stacks of 4-cycle wrapping $D7$ branes allow for a 2nd branch of supersymmetry breaking 
``K\"ahler uplifted" vacua distinct from the LVS branch. 
These vacua can be either AdS, Minkowski or dS by themselves without any need for further `external' sources of uplifting. Their possible existence was first pointed out in~\cite{BB}, while they were shown to be viable in producing controlled large-volume vacua for all moduli in a supergravity analysis~\cite{AW,MRAW}. Recently, \cite{LRVW} provided the first explicit global and consistent F-theory constructions of such K\"ahler uplifted dS vacua. On Swiss-cheese Calabi-Yau threefolds the K\"ahler uplifting branch generates a class of minima for the volume moduli, where $\tau_b\sim N_b$ with $N_b$ the rank of the condensing gauge group of the $D7$ brane stack wrapping the large divisor $D_b$. The blow-up K\"ahler moduli $\tau_i\sim N_i$ are stabilised at smaller volume dictated by the rank of the corresponding condensing gauge groups. For ranks $N_b\sim 
30\div 100$ 
this leads to stabilisation of all K\"ahler moduli at an overall volume ${\cal V}\sim N_b^{3/2}\sim10^{2\div 3}$, with volume moduli masses suppressed by a factor $1/{\cal V}$ compared the masses of the complex structure moduli and the axio-dilaton from fluxes. 
So far, stabilisation of the $h^{1,1}$ $\tau_i$ moduli utilises an identical number of non-perturbative contributions to the superpotential from gaugino condensation. Hence, all associated $c_i$ axions are rendered massive with mass scales tied to 
their scalar partners. K\"ahler uplifting with an (partially) ample divisor utilising less than $h^{1,1}$ instanton effects for stabilisation and in turn realising an axiverse represents an open question.

\end{itemize}

\subsection{Axion and ALP decay constants and their couplings to visible sector particles in the LVS}
\label{AxionDC}

Let us know for the remainder of this review concentrate on the LVS, because, as we have seen, 
it predicts the existence of a QCD axion candidate plus at least one ALP. Their 
decay constants $f_{a_i}$ and their couplings $C_{ij}$ to gauge bosons (or matter fields) $j$ 
can be read off from the matching of the prediction~\eqref{EQ:BIGAXIONCOUPLINGS}  
with the generalisation of the generic low energy effective Lagrangian~\eqref{axion_leff} 
to the case of many axion-like fields, 
by transforming in the former from the original basis $\{c_i\}$ to a basis of fields $\{a_i\}$ 
with canonically normalised kinetic terms~\cite{Cicoli:2012sz}. For the simple 
Swiss-cheese Calabi-Yau setup discussed above and illustrated in Fig.~\ref{fig:lvs}, 
the original axion fields $c_b$ and $c_{\rm vs}$ can be
written in terms of the canonically normalised fields $a_b$ and $a_{\rm vs}$ as:
\begin{eqnarray}
\frac{c_b}{\tau_b} \simeq   
\mathcal{O}\left(1\right)\,a_b +\mathcal{O}\left(\tau_{\rm vs}^{3/4}\,\mathcal{V}^{-1/2}\right) \,a_{\rm vs}\,, \hspace{6ex}
\frac{c_{\rm vs}}{\tau_{\rm vs}} \simeq \mathcal{O}\left(1\right)\,a_b
+\mathcal{O}\left(\tau_{\rm vs}^{-3/4}\,\mathcal{V}^{1/2}\right) a_{\rm vs}\,,
\end{eqnarray}
leading to 
\begin{eqnarray}
f_{a_b} \simeq \frac{M_{\rm KK}}{4\pi}\,, \qquad
f_{a_{\rm vs}} \simeq \frac{M_s}{\sqrt{4\pi}\tau_{\rm vs}^{1/4}}\,,
\label{fasSC}
\end{eqnarray}
\begin{equation}
C_{bb} \simeq   \mathcal{O}\left(1\right), \qquad
C_{{\rm vs}b}\simeq \mathcal{O}\left(\mathcal{V}^{-1/3}\right), \qquad
C_{b{\rm vs}} \simeq  \mathcal{O}\left(\mathcal{V}^{-2/3}\right), \qquad
C_{{\rm vs\,vs}}\simeq \mathcal{O}\left(1\right)\,. 
\end{equation}

The axion $a_{\rm vs}$ can be identified with the QCD axion. 
A range of values for the proper QCD axion decay constant, spanning 
the classic QCD axion window,
\begin{equation}
f_a\equiv \frac{f_{a_{\rm vs}}}{C_{{\rm vs\,vs}}}\sim M_s \sim  
\sqrt{\frac{M_P\, m_{3/2}}{g_s^{1/2} W_0}}\sim 10^{9\div 12}\,{\rm GeV}\,,
\label{eq:lvsexp}
\end{equation} 
are then possible depending upon the exact numerical coefficients, 
the value of the gravitino mass $m_{3/2}\gtrsim $~TeV, and the tuning of $g_s\lesssim 1$ and $W_0$.

The large cycle ALP $a_b$ has a smaller decay constant $f_{a_b}\sim M_{\rm KK}$, 
but its coupling to the standard Model gauge bosons is completely negligible, 
$C_{b{\rm vs}} \simeq  \mathcal{O}\left(\mathcal{V}^{-2/3}\right)$. 

More ALPs, but with decay constant similar to the one of the QCD axion, 
\begin{equation}
f_{\rm ALP_i}\sim f_{a_{\rm vs}}\sim M_s,
\label{alpdecayconstant}
\end{equation}
and coupling to gauge bosons other than the gluon similar 
to the one of the QCD axion, 
\begin{equation}
C_{\rm ALP_i\,vs}\sim C_{{\rm vs\,vs}}\sim \mathcal{O}(1)
\Rightarrow  g_{i\gamma} \equiv \frac{\alpha}{2\pi f_{a_i}} C_{i\gamma} 
\sim 10^{-15}\div 10^{-11}\ {\rm GeV}^{-1},
\label{alpcoupling}
\end{equation}
are obtained,
if there are more small cycles $D_{\rm vs_i}$ intersecting the visible branes, but without
introducing additional D-term conditions. Their masses are expected to be smaller than the mass of the
QCD axion and to be distributed logarithmically hierarchical~\cite{Cicoli:2012sz},
\begin{eqnarray}
m_{\rm ALP_i} \sim e^{- n \pi\tau_{\rm ALP_i}} \times \left\{ \begin{array}{cl}  M_P,  & 
\text{for $\delta W_{\rm np}$ terms or QCD-like masses}, \\
m_{3/2}, & \text{for $\delta K_{\rm np}$ terms}. \end{array}\right. \,
\end{eqnarray}

These findings have been reproduced also in explicit constructions of LVS examples, exploiting 
concrete Calabi-Yau orientifolds and semi-realistic $D7$-brane and flux setups~\cite{Cicoli:2012sz,Cicoli:pascos}.  
That paper considers also LVS variants with asymmetric fibred Calabi-Yaus and 
sequestered SUSY breaking, allowing for more freedom in $m_{3/2}$ and $m_{\rm soft}$, at the
expense of more fine tuning in $W_0$, and discusses briefly the cosmological evolution of the
different LVS incarnations.  

\section{\label{sec:opportunities}Opportunities to probe the intermediate string scale LVS}

\subsection{Haloscope searches}

\begin{center}
\begin{figure}
\begin{center}
\includegraphics[width=0.6\textwidth]{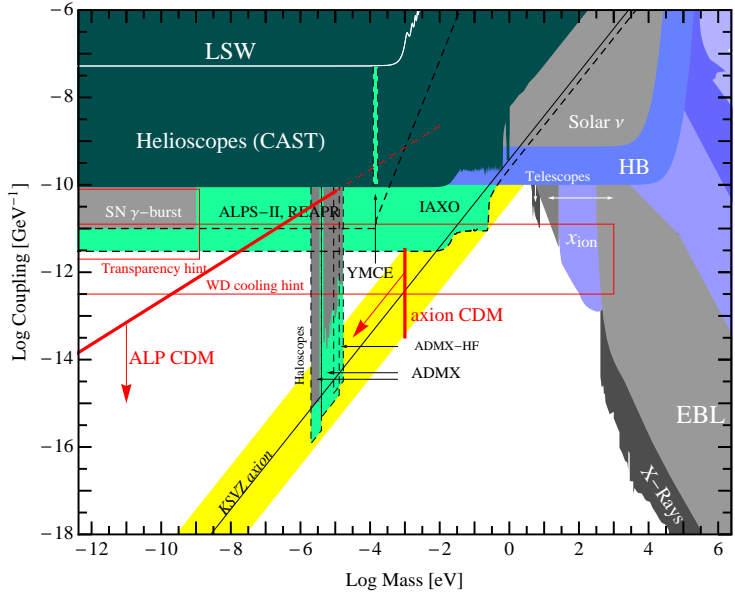} 
\end{center}
\caption{\footnotesize  Axion and ALP coupling to photons, $g_{i \gamma}\equiv \alpha\, C_{i\gamma}/(2\pi f_{a_i})$, vs. its mass (adapted by Javier Redondo~\cite{Redondo:private} from Refs.~\cite{Arias:2012az,Hewett:2012ns}).
The yellow band is the generic prediction for the QCD axion, exploiting Eqs.~\eqref{axionmass} and 
\eqref{axionphotoncoupling}, which relate its mass with its coupling to photons.} 
\label{FIG:Roadmap}
\end{figure}
\end{center}

We have seen, that the LVS predicts -- for the least fine-tuning of fluxes, 
such that $g_s\sim 0.1$ and $W_0\sim 1$, and a TeVish gravitino mass -- an intermediate string scale and thus a 
QCD axion in the classic window, cf. Eq.~\eqref{eq:lvsexp}. For decay constants in the
upper part of this window, $f_a\gtrsim 10^{11\div 12}$~GeV, the QCD axion is expected to 
contribute substantially to the cold dark matter in the universe, see Eq.~\eqref{eq:omegaqcdaxion}.  
Therefore, the intermediate string scale LVS can be probed by haloscope searches for axion cold dark matter~\cite{Sikivie:1983ip} 
such as 
ADMX~\cite{Asztalos:2001tf,Asztalos:2009yp,Asztalos:2011bm,Heilman:2010zz}. These experiments exploit the coupling \eqref{axionphotoncoupling} by 
searching for the signal of dark matter axion to photon conversions in a narrow bandwidth 
microwave cavity sitting in a strong magnetic field.  As can be seen from the light green area in Fig.~\ref{FIG:Roadmap}
labelled as ``Haloscopes", a substantial range of the 
QCD axion dark matter parameter range will be probed by ADMX and other haloscopes~\cite{Baker:2011na} in
the next decade.    

\subsection{Helioscope searches}

A complementary search for the QCD axion in the lower part of the classic window, $f_a\gtrsim 10^{9\div 10}$~GeV,
can be conducted with the next generation of axion helioscopes~\cite{Sikivie:1983ip}, in which one tries to detect
solar axions by their conversion into photons inside of a strong magnet pointed towards the sun. 
Indeed, the projected sensitivity of the proposed International Axion Observatory IAXO~\cite{Irastorza:2011gs} 
covers nicely a part of QCD axion parameter space which will not be covered by the 
haloscope searches, as can be seen in Fig.~\ref{FIG:Roadmap}.    

A very welcome feature of helioscopes is that they do not lose sensitivity towards low masses: 
their projected sensitivity are best and stay constant at small masses, see Fig.~\ref{FIG:Roadmap}. 
That means, with IAXO one may also probe the LVS axiverse, in particular the possible existence of more 
ALPs with approximate the same coupling to photons as the QCD axion. 

This is very important in view of recent tantalising astrophysical hints, such as the anomalous transparency of the Universe for TeV photons~\cite{Horns:2012fx} and the anomalous cooling of white 
dwarfs~\cite{Isern:2008nt,Isern:2008fs}, which could be explained by the existence of an ALP 
with decay constants, couplings and 
mass~\cite{DeAngelis:2007dy,Simet:2007sa,SanchezConde:2009wu,Dominguez:2011xy,Tavecchio:2012um,Isern:2012ef,Horns:2012kw}:  
\begin{equation}
\frac{f_{a_i}}{C_{i\,e}}\simeq (0.7\div 2.6)\times 10^9\ {\rm GeV}\,,
\quad \frac{f_{a_i}}{C_{i\,\gamma}} \sim 10^8\ {\rm GeV},
\quad m_{a_i} \lesssim 10^{-9} \div 10^{-10}\ {\rm eV}\,, 
\label{astrohints}
\end{equation}
compatible with the prediction of an intermediate string scale LVS axiverse with $C_{i\,\gamma}/C_{i\,e} \sim 10$. 
The projected sensitivity of IAXO nicely overlaps with the ALP parameter region required to
explain these hints, see Fig.~\ref{FIG:Roadmap}.
 
\subsection{Light-shining-through-walls searches} 

Intriguingly, this parameter region for ALPs can also be partially probed by purely laboratory based 
light-shining-through-walls experiments~\cite{Redondo:2010dp}, where laser photons are send along 
a strong magnetic field, allowing for their conversion into ALPs, which may then reconvert  
in the strong magnetic field behind a blocking wall into photons, apparently shining through the
wall and susceptible to detection. The projected sensitivities of the proposed experiments 
ALPS-II at DESY and REAPR at Fermilab partially cover the expectations \eqref{alpcoupling} from an intermediate string scale
LVS axiverse and from the hints \eqref{astrohints} from astrophysics, see Figs.~\ref{FIG:Constraints} and \ref{FIG:Roadmap}.

\section{Conclusions}

String phenomenology holds the promise of an axiverse -- the QCD axion plus a (possibly large) number of
further ultralight axion-like particles, possibly populating each decade of mass down to the Hubble scale, $10^{-33}$~eV.
However, although a plenitude of axion-like fields is a generic prediction of string theory, there may be few or no light axions remaining once constraints such as tadpole cancellation and moduli stabilisation are taken into account. 

Interestingly, the promise of an axiverse seems to be fulfilled in the LARGE Volume Scenario (LVS) of IIB string 
flux compactifications. In fact, the simplest globally consistent LVS constructions 
with magnetised $D7$-branes and chirality have at least two light axions: a QCD axion candidate,
with a decay constant of order the string scale, which is intermediate, $f_a\sim M_s\sim M_P/\sqrt{\mathcal{V}}
\sim (M_P m_{3/2}/W_0)^{1/2}\sim 10^{9\div 12}$~GeV, for a TeV scale gravitino mass $m_{3/2}$ and an expectation value
of the flux induced tree-level expectation value $W_0$ of the superpotential of order one, plus 
a nearly decoupled superlight axion-like particle. In setups where the small size branes describing the
visible sector are wrapping more than the minimally required two intersecting four-cycles, 
there are more ultralight axion-like particles which
have approximately the same decay constant and coupling to the photon as the QCD axion candidate. 

At both ends of the above range 
of the decay constant there are exciting phenomenological opportunities. For $f_a\sim 10^{11\div 12}$~GeV, 
the QCD axion can be the dominant part of dark matter and be detected in haloscopes exploiting microwave cavities. 
For $f_a\sim 10^{9\div 10}$~GeV,
the additional ALPs could explain astrophysical anomalies and
be searched for in the upcoming generation
of helioscopes or light-shining-through-a-wall experiments.

\ack 

I would like to thank Michele Cicoli and Mark Goodsell for the great collaboration
on the topics in this review and Alexander Westphal for many enlightning discussions.


\section*{References}

\begin{thebibliography}{9}

  
\bibitem{Weinberg:1977ma} 
  Weinberg S 
  1978 {\em Phys.\ Rev.\ Lett.}\  {\bf 40} 223 

\bibitem{Wilczek:1977pj} 
  Wilczek F 
  1978 {\em Phys.\ Rev.\ Lett.}\  {\bf 40} 279 

\bibitem{Peccei:1977hh} 
  Peccei R D and Quinn H R 
  1977 {\em Phys.\ Rev.\ Lett.}\  {\bf 38} 1440 
  
\bibitem{Georgi:1986df} 
  Georgi H, Kaplan D B and Randall L
  1986 {\em Phys.\ Lett.}\ B {\bf 169} 73 
  
\bibitem{Kim:1979if} 
  Kim J E 
  1979 {\em Phys.\ Rev.\ Lett.}\  {\bf 43} 103 
  
\bibitem{Dine:1981rt} 
  Dine M, Fischler W and Srednicki M 
  1981 {\em Phys.\ Lett.}\ B {\bf 104} 199 
  
\bibitem{Shifman:1979if} 
  Shifman M A , Vainshtein A I and Zakharov V I 
  1980 {\em Nucl.\ Phys.}\ B {\bf 166} 493 
  
\bibitem{Zhitnitsky:1980tq} 
  Zhitnitsky A R 
  1980 {\em Sov.\ J.\ Nucl.\ Phys.}\  {\bf 31} 260 
  [1980 {\em Yad.\ Fiz.}\  {\bf 31} 497]

\bibitem{Bardeen:1977bd} 
  Bardeen W A and Tye S-H H
  1978 {\em Phys.\ Lett.}\ B {\bf 74} 229 
  
\bibitem{Kaplan:1985dv} 
  Kaplan D B 
  1985 {\em Nucl.\ Phys.}\ B {\bf 260} 215 
  
\bibitem{Srednicki:1985xd} 
  Srednicki M 
  1985 {\em Nucl.\ Phys.}\ B {\bf 260} 689 

  
 
 \bibitem{Preskill:1982cy}
  Preskill J, Wise M B and Wilczek F 
  1983 {\em Phys.\ Lett.}\  B {\bf 120} 127

\bibitem{Abbott:1982af}
  Abbott L F and Sikivie P 
  1983 {\em Phys.\ Lett.}\  B {\bf 120} 133 

\bibitem{Dine:1982ah}
  Dine M and Fischler W 
  1983 {\em Phys.\ Lett.}\  B {\bf 120} 137
 
  
\bibitem{Witten:1984dg} 
  Witten E 
  1984 {\em Phys.\ Lett.}\ B {\bf 149} 351 
  
\bibitem{Conlon:2006tq} 
  Conlon J P 
  2006 {\em JHEP} {\bf 0605} 078 

\bibitem{Svrcek:2006yi} 
  Svrcek P and Witten E 
  2006 {\em JHEP} {\bf 0606} 051 
  
\bibitem{Conlon:2006ur} 
  Conlon J P 
  2006 {\em Phys.\ Rev.\ Lett.}\  {\bf 97} 261802 

\bibitem{Choi:2006qj} 
  Choi K-S, Kim I-W and Kim J E
  2007 {\em JHEP} {\bf 0703} 116 

\bibitem{Arvanitaki:2009fg} 
  Arvanitaki A {\em et al}, 
  2010 {\em Phys.\ Rev.}\ D {\bf 81} 123530 

\bibitem{Acharya:2010zx} 
  Acharya B S, Bobkov K and Kumar P 
  2010 {\em JHEP} {\bf 1011} 105 

\bibitem{Cicoli:2012sz} 
  Cicoli M, Goodsell M and Ringwald A
  2012 {\em Preprint} arXiv:1206.0819 [hep-th]
  
\bibitem{Grimm:2004uq}
  Grimm T W and Louis J
  2004 {\em Nucl.\ Phys.}\ B {\bf 699} 387

\bibitem{Jockers:2004yj}
  Jockers H and Louis J 
  2005 {\em Nucl.\ Phys.}\  B {\bf 705} 167

\bibitem{Goodsell:2009xc}
  Goodsell M, Jaeckel J, Redondo J and Ringwald A
  2009 {\em JHEP} {\bf 0911} 027
  
\bibitem{Cicoli:2011yh}
  Cicoli M, Goodsell M, Jaeckel J and Ringwald A 
  2011 {\em JHEP} {\bf 1107} 114

\bibitem{Banks:2003es}
  Banks T, Dine M and Gorbatov E
  2004 {\em JHEP} {\bf 0408}  058
  
\bibitem{Donoghue:2003vs}
  Donoghue J F 
  2004 {\em Phys.\ Rev.}\ D {\bf 69}  106012
   [2004 {\em Erratum-ibid.}\ D {\bf 69} 129901]

\bibitem{GVW}
Gukov S, Vafa C and Witten E
  2000 {\em Nucl.\ Phys.}\ B {\bf 584 }   69

\bibitem{gkp}
Giddings S B, Kachru S and Polchinski J
  2002 {\em Phys.\ Rev.}\  D {\bf 66} 106006

\bibitem{kklt}
  Kachru S, Kallosh R, Linde A and Trivedi S P
  2003 {\em Phys.\ Rev.}\  D {\bf 68} 046005

\bibitem{Bobkov:2010rf}
  Bobkov K, Braun V, Kumar P and Raby S
  2010 {\em JHEP} {\bf 1012}  056

\bibitem{Balasubramanian:2005zx}
  Balasubramanian V, Berglund P, Conlon J P and Quevedo F 
  2005 {\em JHEP} {\bf 0503}  007

  
\bibitem{Cicoli:2011qg}
  Cicoli M, Mayrhofer C and Valandro R
  2012 {\em JHEP} {\bf 1202} 062
  
\bibitem{Cicoli:pascos}
  Cicoli M 
  2012 these proceedings  
 

\bibitem{LVSspectrum}
  Conlon J P, Quevedo F and Suruliz K 
  2005 {\em JHEP} {\bf 0508} 007  

\bibitem{alphaprime}
  Becker K, Becker M, Haack M and Louis J 
  2002 {\em JHEP} {\bf 0206} 060  

\bibitem{ExplicitSC}
  Collinucci A {\em et al} 
  2009 {\em JHEP} {\bf 0907}  074


\bibitem{BB}
  Balasubramanian V and Berglund P
  2004 {\em JHEP} {\bf 0411} 085 

\bibitem{AW} 
  Westphal A 
  2007 {\em JHEP} {\bf 0703} 102 

\bibitem{MRAW}
  Rummel M and Westphal A 
  2012 {\em JHEP} {\bf 1201} 020 

\bibitem{LRVW}
  Louis J, Rummel M, Valandro R and Westphal A 
  {\em Preprint} arXiv:1208.3208 [hep-th]
 

\bibitem{Sikivie:1983ip}
  Sikivie P 
  2983 {\em Phys.\ Rev.\ Lett.}\  {\bf 51} 1415
  [1984 {\em Erratum-ibid.}\  {\bf 52} 695].

\bibitem{Asztalos:2001tf}
  Asztalos S J {\em et al} 
  2001 {\em Phys.\ Rev.}\ D {\bf 64}  092003

\bibitem{Asztalos:2009yp}
  Asztalos S J {\it et al}  [The ADMX Collaboration] 
  2010 {\em Phys.\ Rev.\ Lett.}\  {\bf 104}  041301
  
 \bibitem{Asztalos:2011bm}
  Asztalos S J {\em et al} 
  2011 {\em Nucl.\ Instrum.\ Meth.}\ A {\bf 656} 39

\bibitem{Heilman:2010zz}
  Heilman J {\it et al}  [ADMX Collaboration] 
  2010 {\em AIP Conf.\ Proc.}\  {\bf 1274}  115 
 
\bibitem{Baker:2011na}
  Baker O K {\it et al} 
  2012 {\em Phys.\ Rev.}\ D {\bf 85}  035018


\bibitem{Redondo:private} 
  Redondo J 2012 private communication

\bibitem{Arias:2012az}
  Arias P, Cadamuro D, Goodsell M, Jaeckel J, Redondo J and Ringwald A 
  2012 {\em JCAP} {\bf 1206}  013

\bibitem{Hewett:2012ns} 
  Hewett J L {\it et al.} 
  2012 {\em Preprint} arXiv:1205.2671 [hep-ex]

\bibitem{Irastorza:2011gs}
  Irastorza I G {\em et al} 
  2011 {\em JCAP} {\bf 1106}  013

\bibitem{Horns:2012fx}
  Horns D and Meyer M 
  2012 {\em JCAP} {\bf 1202}  033

\bibitem{Isern:2008nt}
  Isern J, Garcia-Berro E, Torres S and Catalan S
  2008 {\em Astrophys.\ J.}\  {\bf 682} L109

\bibitem{Isern:2008fs}
  Isern J, Catalan S, Garcia-Berro E and Torres S 
  2009 {\em J.\ Phys.\ Conf.\ Ser.}\  {\bf 172}  012005

\bibitem{DeAngelis:2007dy}
  De Angelis A, Mansutti O, Roncadelli M 
  2007 {\em Phys.\ Rev.}\  {\bf D76 }   121301

\bibitem{Simet:2007sa}
  Simet M, Hooper D, Serpico P D 
  2008 {\em Phys.\ Rev.}\  {\bf D77 }  063001 

\bibitem{SanchezConde:2009wu}
  Sanchez-Conde M A {\em et al} 
  2009 {\em Phys.\ Rev.}\  {\bf D79 }  123511 


\bibitem{Dominguez:2011xy}
  Dominguez A, Sanchez-Conde M A and Prada F 
  2011 {\em JCAP} {\bf 1111} 020

\bibitem{Tavecchio:2012um}
  Tavecchio F, Roncadelli M, Galanti G and Bonnoli G 
  2012 {\em Preprint} arXiv:1202.6529 [astro-ph.HE]
  

\bibitem{Isern:2012ef}
  Isern J {\em et al}, 
  2012 {\em Preprint} arXiv:1204.3565 [astro-ph.SR] 

\bibitem{Horns:2012kw}
  Horns D {\em et al} 
  2012 {\em Preprint} arXiv:1207.0776 [astro-ph.HE]

 
\bibitem{Redondo:2010dp}
  Redondo J and Ringwald A 
  2011 {\em Contemp.\ Phys.}\  {\bf 52}  211
  

\end{thebibliography}
\end{document}